\documentclass{article}
\usepackage{amsmath}
\usepackage{amsfonts}
\usepackage{amssymb}
\usepackage{graphicx}
\begin{document}

\title{The Entropic Dynamics of Relativistic \\Quantum Fields\thanks{Presented at MaxEnt 2012, The 32nd
International Workshop on Bayesian Inference and Maximum Entropy Methods in
Science and Engineering, (July 15--20, 2012, Garching, Germany). } }
\author{Ariel Caticha\\{\small Department of Physics, University at Albany-SUNY, }\\{\small Albany, NY 12222, USA.}}
\date{}
\maketitle

\begin{abstract}
The formulation of quantum mechanics within the framework of entropic dynamics
is extended to the domain of relativistic quantum fields. The result is a
non-dissipative relativistic diffusion in the $\infty$-dimensional space of
field configurations. On extending the notion of entropic time to the
relativistic regime we find that the field fluctuations provide the clock that
sets the scale of duration. We also find that the usual divergences that
affect all quantum field theories do not refer to the real values of physical
quantities but rather to epistemic quantities invariably associated to
unphysical probability distributions such as variances and other measures of uncertainty.
\end{abstract}

\section{Introduction}

The goal of the Entropic Dynamics (ED) framework is to seek for quantum
mechanics a level of understanding comparable to that attained by Jaynes in
statistical mechanics and thermodynamics. The challenge is graphically
described by Jaynes' omelette metaphor: \textquotedblleft Our present QM
formalism is a peculiar mixture describing in part realities in Nature, in
part incomplete human information about Nature --- all scrambled up by
Heisenberg and Bohr into an omelette that nobody has seen how to
unscramble.\textquotedblright\cite{Jaynes 1990}\ 

In previous work the Schr\"{o}dinger equation for the non-relativistic quantum
dynamics of $N$ particles was derived as an example of entropic
inference.\cite{Caticha 2010a} Within the ED framework the ontic and the
epistemic elements of the model are neatly unscrambled.\footnote{The
distinction ontic/epistemic is not to be confused with the distinction
objective/subjective: a probability is a purely epistemic notion that
incorporates both subjective and objective elements. Indeed, on one hand, the
assignment of priors and likelihoods involves judgments that are inevitably
subjective and, on the other hand, the very reason why we collect data and
other information is precisely in order to update probabilities and thereby
enhance their objectivity.} The non-relativistic model proposed in
\cite{Caticha 2010a} has a clear ontology of particles with real and definite,
albeit uncertain, positions. This is in stark contrast with the standard
interpretation of quantum mechanics which refrains from attributing definite
values to any observable prior to an actual measurement. The wave function and
other \textquotedblleft observables\textquotedblright\ such as momentum and
energy do not represent anything physically real. They reflect epistemic
rather than ontic aspects of the model. They are properties associated to the
probability distributions and not to the particles themselves.\cite{Nawaz
Caticha 2011}\cite{Johnson Caticha 2011}

But for such an epistemic view of quantum states to be satisfactory it is not
sufficient to accept that $\left\vert \psi\right\vert ^{2}$ represents a state
of knowledge; all other features of the wave function must be epistemic too.
One must provide an epistemic interpretation for the phase of the wave
function and, furthermore, one must show that changes or updates of the
epistemic $\psi$ --- including both unitary time evolution according to the
Schr\"{o}dinger equation and the projection postulate during measurement ---
occur precisely according to the stipulations of entropic and Bayesian
inference.\cite{Caticha 2012} The ED framework has led to several new insights
including the entropic nature of time \cite{Caticha 2010b}; an entropic
interpretation of the phase of the wave function and of gauge transformations;
the role of Hilbert spaces and complex numbers; the uncertainty relations
\cite{Nawaz Caticha 2011}; and the quantum measurement problem and the
interpretation of \textquotedblleft observables\textquotedblright\ other than
position \cite{Johnson Caticha 2011}.

In this paper we take the first step towards extending entropic dynamics to
the domain of relativistic scalar quantum fields. The basic dynamical
quantity, the probability of a small change, is found maximizing the
appropriate entropy subject to suitable constraints. Then the notion of time
is introduced to keep track of the accumulation of small changes. The
resulting evolution is described by a functional Fokker-Planck equation and a
(quantum and relativistic) functional Hamilton-Jacobi equation. These
equations are then combined into a functional Schr\"{o}dinger equation which
is formally equivalent to the standard quantum theory of scalar fields (see
\emph{e.g.}, \cite{Jackiw 1989}).

Many of the standard predictions follow immediately. For example, the excited
states of the field can be interpreted as identical particles that obey
Bose-Einstein statistics. We also find the usual infinities that plague all
quantum field theories but with one major difference: the divergences do not
refer to the real values of physical quantities but rather to unphysical
epistemic quantities such as variances and other measures of uncertainty.
Thus, in entropic dynamics the infinities of quantum field theory are not
physical effects; they are indications that the information that has been
included in the analysis is insufficient to answer certain questions. This
result supports Jaynes' intuition foreseeing \emph{\textquotedblleft...the
possibility of a future quantum theory in which the role of incomplete
information is recognized ... when we free ourselves from the delusion that
probabilities are physically real things, then when }[a dispersion]\emph{\ }%
$\Delta F$\emph{\ is infinite, that does not mean that any physical quantity
is infinite. It means only that the theory is completely unable to predict
}$F$\emph{. The only thing that is infinite is the uncertainty of the
prediction.\textquotedblright}\cite{Jaynes 1990} \emph{\ }

\section{Entropic dynamics}

We are concerned with the dynamics of a single scalar field. The
generalization to other Boson fields is immediate. A particular field
configuration $\phi(x)$ associates one degree of freedom to each spatial point
$x$ in three dimensional Euclidean space. Such a field is represented as a
point $\phi\in\mathcal{F}$ in the $\infty$-dimensional configuration space
$\mathcal{F}$. Our first assumption is that the space $\mathcal{F}$ is flat
and its metric is a straightforward generalization of the metric $\delta_{ij}$
of Euclidean space so that the distance $\Delta\ell$ between two slightly
different configurations $\phi$ and $\phi+\Delta\phi$ is written as
\begin{equation}
\Delta\ell^{2}=%
{\textstyle\int}
d^{3}x\,[\Delta\phi(x)]^{2}~. \label{distance}%
\end{equation}
The justification of this assumption and of several others that will follow
is, in the end, purely pragmatic. An analogy can be drawn to other physical
theories. For example, in Newtonian mechanics different physical situations
are described by different forces; in an entropic framework different physical
situations are described by different constraints. (We shall later see that
$\Delta\ell$ will enter as a constraint.) And just as Newtonian mechanics did
not justify the $1/r^{2}$ force law except through its empirical success, we
will not, at this early point in the development of the ED of quantum fields,
offer any deeper insight into the choice of distance except to note that it is
empirically successful too.

The second assumption is that in addition to the field $\phi$ the world
contains other stuff described by variables $y$ living in some space
$\mathcal{Y}$.\footnote{In statistics such variables are called
\emph{nuisance} variables. Although we are not directly interested in them
they affect the variables we do care about and must be included in the
analysis.} The number and nature of the extra variables $y\in\mathcal{Y}$ need
not be specified; we only need to assume that their values are uncertain and
that this uncertainty is described by some probability distribution
$p[y|\phi]$ that depends on the particular field configuration $\phi
$.\footnote{The notation we adopt is standard in theoretical physics: the
$x$-dependence is denoted as a subscript, $\phi(x)=\phi_{x}$; square brackets
as in $F[\phi]$ denote functionals; functional derivatives are written
$\delta/\delta\phi_{x}$; and functional integrals are written as $%
{\textstyle\int}
D\phi\,F[\phi]$.}

The $\infty$-dimensional manifold of distributions $p[y|\phi]$ --- for each
field configuration $\phi$ there is a corresponding $p[y|\phi]$ --- is a
\emph{statistical manifold} $\mathcal{M}$ and each distribution $p[y|\phi
]\in\mathcal{M}$ can be conveniently labeled by its corresponding $\phi$. For
future reference, the entropy $S[\phi]$ of $p[y|\phi]$ relative to an
underlying measure $q[y]$ of the space $\mathcal{Y}$ is given by the
functional integral
\begin{equation}
S[\phi]=-\int Dy\,p[y|\phi]\log\frac{p[y|\phi]}{q[y]}~. \label{entropy a}%
\end{equation}

The dynamics follows from yet a third assumption, that large changes result
from the accumulation of many small changes. Thus, the basic dynamical problem
is to find the transition probability $P[\phi^{\prime}|\phi]$ of a small
change from an initial $\phi$ to a new $\phi^{\prime}=\phi+\Delta\phi$. Since
neither the new field $\phi^{\prime}$ nor the new $y^{\prime}$ are known the
relevant space is $\mathcal{F\times Y}$ and we seek the joint distribution
$P[\phi^{\prime},y^{\prime}|\phi]$. This is found maximizing the (relative)
entropy,
\begin{equation}
\mathcal{S}[P,Q]=-\int D\phi^{\prime}Dy^{\prime}\,P[\phi^{\prime},y^{\prime
}|\phi]\log\frac{P[\phi^{\prime},y^{\prime}|\phi]}{Q[\phi^{\prime},y^{\prime
}|\phi]}~, \label{Sppi}%
\end{equation}
subject to constraints that codify the appropriate relevant information.

\noindent\textbf{The prior:} We assume a state of extreme ignorance
represented by a product,
\begin{equation}
Q[\phi^{\prime},y^{\prime}|\phi]=q[\phi^{\prime}]q[y^{\prime}]~, \label{prior}%
\end{equation}
where $q[\phi]\ $and $q[y]$ are uniform distributions.

\noindent\textbf{The first constraint:} Use the product rule to write
$P[\phi^{\prime},y^{\prime}|\phi]=P[\phi^{\prime}|\phi]P[y^{\prime}%
|\phi^{\prime},\phi]$. The first factor $P[\phi^{\prime}|\phi]$ is the
transition probability we want to find. The second factor, $P[y^{\prime}%
|\phi^{\prime},\phi]$, is constrained to remain on the manifold $\mathcal{M}$,
that is $P[y^{\prime}|\phi^{\prime},\phi]=p[y^{\prime}|\phi^{\prime}%
]\in\mathcal{M}$. Therefore,
\begin{equation}
P[\phi^{\prime},y^{\prime}|\phi]=P[\phi^{\prime}|\phi]p[y^{\prime}%
|\phi^{\prime}]~. \label{constraint p}%
\end{equation}
This constraint is implemented by direct substitution into the entropy
(\ref{Sppi}):
\begin{equation}
\mathcal{S}[P,Q]=-\int D\phi^{\prime}\,P[\phi^{\prime}|\phi]\log\frac
{P[\phi^{\prime}|\phi]}{q[\phi^{\prime}]}+\int D\phi^{\prime}\,P[\phi^{\prime
}|\phi]\,S[\phi^{\prime}]~
\end{equation}
where $S[\phi^{\prime}]$ is given by eq.(\ref{entropy a}).

\noindent\textbf{The second constraint }represents the fact that physical
changes are not discontinuous: the requirement that $\phi^{\prime}$ be an
infinitesimally close to $\phi$ is implemented by imposing that the
expectation
\begin{equation}
\left\langle \Delta\ell^{2}\right\rangle =\int D\phi^{\prime}\,P[\phi^{\prime
}|\phi]\int d^{3}x\,(\Delta\phi_{x})^{2}=\Delta\lambda^{2}[\phi]~
\label{short step}%
\end{equation}
take some infinitesimal but for now unspecified value $\Delta\lambda^{2}$
which could depend on $\phi$.

Maximizing $\mathcal{S}$ subject to the constraints above plus normalization
yields,
\begin{equation}
P[\phi^{\prime}|\phi]=\frac{1}{\zeta}\exp\left(  S[\phi^{\prime}]-\frac{1}%
{2}\alpha\lbrack\phi]\int d^{3}x\,(\phi_{x}^{\prime}-\phi_{x})^{2}\right)  ~,
\label{tr prob a}%
\end{equation}
where $\zeta$ is a normalization constant and $\alpha\lbrack\phi]$ is the
Lagrange multiplier that implements the constraint (\ref{short step}). For
small $\Delta\phi$ which corresponds to large $\alpha$, we can expand%
\begin{equation}
S[\phi+\Delta\phi]=S[\phi]+\int d^{3}x\,\frac{\delta S}{\delta\phi_{x}}%
\Delta\phi_{x}~, \label{delta S}%
\end{equation}
and substitute into (\ref{tr prob a}) to get (after some algebra; see
\cite{Caticha 2012})%
\begin{equation}
P[\phi^{\prime}|\phi]=\frac{1}{Z}\exp\left(  -\frac{1}{2}\alpha\lbrack
\phi]\int d^{3}x\,(\Delta\phi_{x}-\Delta\bar{\phi}_{x})^{2}\right)  ~,
\label{tr prob b}%
\end{equation}
where $Z$ is a new normalization constant and
\begin{equation}
\Delta\bar{\phi}_{x}=\langle\Delta\phi_{x}\rangle=\frac{1}{\alpha}\frac{\delta
S}{\delta\phi_{x}}~. \label{drift}%
\end{equation}
The transition probability given by (\ref{tr prob b}) and (\ref{drift}) is the
basis for dynamics: $\phi_{x}$ changes by a small amount $\Delta\phi
_{x}=\Delta\bar{\phi}_{x}+\Delta w_{x}$ given by a drift $\Delta\bar{\phi}%
_{x}$ plus a fluctuation $\Delta w_{x}$ such that
\begin{equation}
\langle\Delta w_{x}\rangle=0\quad\text{and}\quad\langle\Delta w_{x}\Delta
w_{x^{\prime}}\rangle=\frac{1}{\alpha\lbrack\phi]}\delta_{xx^{\prime}}~.
\label{fluct a}%
\end{equation}
For large $\alpha$ the fluctuations $\Delta w\sim O(\alpha^{-1/2})$ dominate
over the drift $\Delta\bar{\phi}\sim O(\alpha^{-1})$ which means that the
trajectory in configuration space $\mathcal{F}$ is continuous but
non-differentiable --- a Brownian motion. The choice of $\Delta\lambda
^{2}[\phi]$ or equivalently of its multiplier $\alpha\lbrack\phi]$ is fixed by
a symmetry argument. For infinitesimal $\Delta\phi$ the dynamics is dominated
by the fluctuations $\Delta w$. To reflect the translational symmetry of the
flat configuration space $\mathcal{F}$ we choose $\alpha\lbrack\phi]$ so that
the fluctuations in eq.(\ref{fluct a}) be independent of $\phi$. Therefore
$\alpha\lbrack\phi]=\operatorname{constant}$.

\section{Accumulating changes: entropic time}

Time is intimately related to change. In ED time is introduced as a convenient
book-keeping device to keep track of the accumulation of small changes. The
basic strategy, described in detail in \cite{Caticha 2010a}\cite{Caticha
2010b}\cite{Caticha 2012}, is to develop a model that includes (a) something
one might identify as an \textquotedblleft instant\textquotedblright, (b) a
sense in which these instants can be \textquotedblleft
ordered\textquotedblright, (c) a convenient concept of \textquotedblleft
duration\textquotedblright\ measuring the separation between instants. This
set of concepts constitutes what we will call \textquotedblleft entropic
time\textquotedblright. Incidentally, the model incorporates an intrinsic
directionality --- an evolution from past instants towards future instants.
Thus, an arrow of time is generated automatically.\cite{Caticha 2010b}

When referring to the probability $\rho\lbrack\phi]$ of a particular field
configuration $\phi$ it is implicit that the values $\phi_{x}$ at different
locations $x$ occur at the same instant. In ED we turn this intuition around
and use it to define the notion of instant: an instant $t$ is defined by a
probability distribution $\rho_{t}[\phi]$. Such instants are naturally ordered
by the dynamics: if the distribution $\rho_{t}[\phi]$\ refers to a certain
instant $t$, and $P[\phi^{\prime}|\phi]$ in (\ref{tr prob b}) is the
probability of a small change to $\phi^{\prime}=\phi+\Delta\phi$, then we can
construct the distribution%

\begin{equation}
\rho_{t^{\prime}}[\phi^{\prime}]=\int D\phi\,P[\phi^{\prime}|\phi]\,\rho
_{t}[\phi]~, \label{CK}%
\end{equation}
and use it to define what we mean by the \textquotedblleft
next\textquotedblright\ instant, $t^{\prime}=t+\Delta t$. \noindent Thus,
eq.(\ref{CK}) allows entropic time to be constructed, step by step, as a
succession of instants.\emph{\ }Finally, to establish a measure of duration
between successive instants we consult the dynamics again --- \emph{time is
defined so that motion looks simple. }We define the multiplier $\alpha(t)$ to
be independent of $t$,%

\begin{equation}
\alpha(t)=\frac{1}{\eta\Delta t}=\operatorname{constant}~, \label{alpha}%
\end{equation}
where $\eta$ is a constant introduced so that $\Delta t$ has units of time.
(It is further convenient to choose units so that the speed of light $c=1$.)
Thus, the drift velocity is
\begin{equation}
b_{x}[\phi]=\frac{\Delta\bar{\phi}_{x}}{\Delta t}=\eta\frac{\delta S[\phi
]}{\delta\phi_{x}}~, \label{drift velocity}%
\end{equation}
and the fluctuations are given by
\begin{equation}
\langle\Delta w_{x}\rangle=0\quad\text{and}\quad\langle\Delta w_{x}\Delta
w_{x^{\prime}}\rangle=\langle\Delta\phi_{x}\Delta\phi_{x^{\prime}}\rangle
=\eta\Delta t\,\delta_{xx^{\prime}}~. \label{fluct b}%
\end{equation}
With this choice of $\alpha$ the strength of the fluctuations remains constant
in time. Or, in other words: \emph{the Brownian field fluctuations constitute
the standard clock that sets the scale of entropic time}.

We are now ready to study how small changes $\Delta\phi$ accumulate as
eq.(\ref{CK}) is iterated. The result, well known from diffusion theory (see
\emph{e.g. }\cite{Caticha 2012} for details), is a functional Fokker-Planck
equation which can be written as a continuity equation,%
\begin{equation}
\partial_{t}\rho_{t}[\phi]=-\int d^{3}x\frac{\delta}{\delta\phi_{x}}\left(
\rho_{t}[\phi]v_{x}[\phi]\right)  ~, \label{FP b}%
\end{equation}
where $v_{x}[\phi]$ is the velocity of the probability flow in the
$\mathcal{F}$ space or \emph{current velocity},
\begin{equation}
v_{x}[\phi]=b_{x}[\phi]+u_{x}[\phi]~,
\end{equation}
and $u_{x}[\phi]$ is the \emph{osmotic velocity}
\begin{equation}
u_{x}[\phi]=-\eta\frac{\delta\log\rho_{t}^{1/2}}{\delta\phi_{x}}~.
\label{osmo}%
\end{equation}
The osmotic contribution to the probability flow, $\rho_{t}u_{x}$, is the
diffusion current in $\mathcal{F}$ space. Since both the drift velocity
$b_{x}$ and the osmotic velocity $u_{x}$ are gradients in $\mathcal{F}$ space,
it follows that the current velocity is a gradient too,
\begin{equation}
v_{x}[\phi]=\eta\frac{\delta\Phi\lbrack\phi]}{\delta\phi_{x}}\quad
\text{where}\quad\Phi\lbrack\phi]=S[\phi]-\log\rho_{t}^{1/2}[\phi]~.
\label{curr}%
\end{equation}

\section{Non-dissipative diffusion}

The implicit constraint that the statistical manifold $\mathcal{M}$ is rigidly
fixed has led us to describe the evolution of $\rho_{t}[\phi]$ as a diffusion
process in $\mathcal{F}$ space but quantum mechanics is not just diffusion. We
will therefore modify this constraint by allowing $\mathcal{M}$ to participate
in the dynamics, that is, the distribution $p_{t}[y|\phi]$, its entropy
$S_{t}[\phi]$, and the \textquotedblleft phase\textquotedblright\ functional,%

\begin{equation}
\Phi_{t}[\phi]=S_{t}[\phi]-\log\rho_{t}^{1/2}[\phi]~, \label{phase}%
\end{equation}
all become time-dependent. The dynamics of $\mathcal{M}$ is specified by
imposing the conservation of a certain functional $E[\rho_{t},S_{t}]$ of
probability and entropy that we will call the \textquotedblleft
energy\textquotedblright. Note that \emph{this \textquotedblleft
energy\textquotedblright\ is an epistemic concept: it is a property not of the
physical field but of our unphysical state of knowledge}. The proposed
\textquotedblleft energy\textquotedblright\ functional is chosen to be the
expectation of a local density,%
\begin{equation}
E[\rho_{t},\Phi_{t}]=\int D\phi\,\rho_{t}[\phi]\int d^{3}x\,\mathcal{E}%
(\rho_{t},\partial\rho_{t},\Phi_{t},\partial\Phi_{t})~. \label{energy a}%
\end{equation}
The density $\mathcal{E}$ is chosen so that it is invariant under time
reversal and consists of the lowest non-trivial powers of the current and
osmotic velocities,%
\begin{equation}
\mathcal{E}(\rho_{t},\partial\rho_{t},\Phi_{t},\partial\Phi_{t})=\frac
{\eta^{2}}{2}\left(  \frac{\delta\Phi\lbrack\phi]}{\delta\phi_{x}}\right)
^{2}+a\frac{\eta^{2}}{2}\left(  \frac{\delta\log\rho_{t}^{1/2}}{\delta\phi
_{x}}\right)  ^{2}+V(\phi_{x},\partial\phi_{x})~. \label{energy b}%
\end{equation}
The first term $v_{x}^{2}/2$ represents \textquotedblleft
kinetic\textquotedblright\ energy.\footnote{An overall multiplicative constant
has been adjusted so the coefficient of the kinetic energy is $1/2$.
\par
{}} The second term $au_{x}^{2}/2$ represents an osmotic \textquotedblleft
potential\textquotedblright\ energy where the constant $a$ measures its
strength relative to the kinetic energy. The last term represents the more
standard contribution to potential energy; in general $V$ will depend on the
field $\phi_{x}$ and its spatial derivatives $\partial\phi_{x}$.

Taking the time derivative of (\ref{energy a}), using eqs.(\ref{FP b}),
(\ref{osmo}) and (\ref{curr}), after integration by parts and some algebra
(eventually) yields%
\begin{equation}
\dot{E}=\int D\phi\,\dot{\rho}_{t}\left[  \eta\dot{\Phi}+%
{\textstyle\int}
d^{3}x\left(  \frac{\eta^{2}}{2}(\frac{\delta\Phi}{\delta\phi_{x}})^{2}%
-a\frac{\eta^{2}}{2}\frac{1}{\rho_{t}^{1/2}}\frac{\delta^{2}\rho_{t}^{1/2}%
}{\delta\phi_{x}^{2}}+V\right)  \right]  \,. \label{energy dot}%
\end{equation}
Now, any instant $t$ can be taken as the initial instant for evolution into
the future. We impose that the energy $E$ be conserved for any arbitrary
choice of initial conditions, namely $\rho_{t}[\phi]$ and $\Phi_{t}[\phi]$,
which implies an arbitrary choice of $\dot{\rho}_{t}$. Therefore,
\begin{equation}
\eta\dot{\Phi}=-\int d^{3}x\left(  \frac{\eta^{2}}{2}(\frac{\delta\Phi}%
{\delta\phi_{x}})^{2}-a\frac{\eta^{2}}{2}\frac{1}{\rho_{t}^{1/2}}\frac
{\delta^{2}\rho_{t}^{1/2}}{\delta\phi_{x}^{2}}+V\right)  ~, \label{SEb}%
\end{equation}
which we recognize as a functional form of the quantum Hamilton-Jacobi equation.

We are done. Equations (\ref{SEb}) and the Fokker-Planck equation
eq.(\ref{FP b}) with (\ref{curr}),
\begin{equation}
\dot{\rho}_{t}=-\eta\int d^{3}x\frac{\delta}{\delta\phi_{x}}\left(  \rho
_{t}\frac{\delta\Phi_{t}}{\delta\phi_{x}}\right)  ~, \label{SEa}%
\end{equation}
are the coupled dynamical equations we seek. They describe energy conservation
and entropic diffusion respectively. Eq.(\ref{SEb}) shows how $\rho_{t}$
affects the evolution of $\Phi_{t}$ and eq.(\ref{SEa}) shows how $\Phi_{t}$
affects the evolution of $\rho_{t}$.

We can always combine the functions $\rho_{t}$ and $\phi_{t}$ into the complex
function $\Psi_{t}[\phi]=\rho_{t}^{1/2}\exp(i\Phi_{t})$, and rewrite
eqs.(\ref{SEb}) and (\ref{SEa}) as a single complex equation,%
\begin{equation}
i\eta\partial_{t}\Psi_{t}=\int d^{3}x\left(  -\frac{\eta^{2}}{2}\frac
{\delta^{2}}{\delta\phi_{x}^{2}}+V+(1-a)\frac{\eta^{2}}{2}\frac{1}{\rho_{t}%
}\frac{\delta^{2}\rho_{t}}{\delta\phi_{x}^{2}}\right)  \Psi_{t}~.
\end{equation}
Next set $\eta=\hbar$ and choose $a=1$ to get the functional Schr\"{o}dinger
equation,\footnote{As discussed in \cite{Caticha 2010a} the choice $a=1$ does
not represent any loss of generality. It can always be attained by an
appropriate rescaling of the units of $\eta=\kappa\eta_{\text{new}}$ and
regraduation of $\Phi=\Phi_{\text{new}}/\kappa$.}
\begin{equation}
i\hbar\partial_{t}\Psi_{t}=\int d^{3}x\left(  -\frac{\hbar^{2}}{2}\frac
{\delta^{2}}{\delta\phi_{x}^{2}}+V(\phi_{x},\partial\phi_{x})\right)  \Psi
_{t}~,
\end{equation}
which concludes the derivation. At this point the potential $V(\phi
_{x},\partial_{x}\phi_{x})$ is essentially arbitrary. A reasonable form is
obtained by doing a Taylor expansion about weak fields and gradients and then
imposing the rotational and Lorentz symmetries required by the experimental evidence,%

\begin{equation}
V(\phi_{x},\partial\phi_{x})=(\partial\phi_{x})^{2}+m^{2}\phi_{x}^{2}%
+\lambda^{\prime}\phi_{x}^{3}+\lambda^{\prime\prime}\phi_{x}^{4}%
+\ldots\label{potential}%
\end{equation}
The various coefficients represent mass and other coupling constants. We
conclude that the ED framework reproduces the standard relativistic quantum
theory of scalar fields.\cite{Jackiw 1989}

\section{Conclusions and discussion}

Entropic dynamics provides an alternative method of quantization --- entropic
quantization. In the ED framework quantum field theory is a non-dissipative
diffusion in the configuration space $\mathcal{F}$.

Entropic time is defined so that motion looks simple. In Newtonian mechanics
free particles provide the clock and time is defined so that free particles
cover equal distances in equal times. In the ED of fields, the field
fluctuations provide the clock and entropic time is defined so that field
fluctuations are uniform in space and time.

The standard predictions of quantum field theory follow immediately --- just
transform from the Schr\"{o}dinger representation to the Heisenberg operator
representation. For example, if we restrict ourselves to the first two terms
in (\ref{potential}) we obtain the Schr\"{o}dinger representation of the free
Klein-Gordon field,
\begin{equation}
i\hbar\partial_{t}\Psi=\frac{1}{2}\int d^{3}x\left(  -\hbar^{2}\frac
{\delta^{2}}{\delta\phi_{x}^{2}}+(\partial\phi_{x})^{2}+m^{2}\phi_{x}%
^{2}\right)  \Psi~.
\end{equation}
A standard calculation \cite{Long Shore 1996} of the ground state yields,
\begin{equation}
\Psi_{t}^{(0)}[\phi]=e^{-iE_{0}t/\hbar}\exp-\frac{1}{2}\int d^{3}%
xd^{3}x^{\prime}\phi(\vec{x})G(\vec{x},\vec{x}^{\prime})\phi(\vec{x}^{\prime})
\end{equation}
where the ground state energy, $E_{0}=\langle%
{\textstyle\int}
d^{3}x\,\mathcal{E}\rangle_{0}$, is divergent:%
\[
E_{0}=\frac{1}{2}\int d^{3}x\,G(\vec{x},\vec{x})\quad\text{where}\quad
G(\vec{x},\vec{x}^{\prime})=\int\frac{d^{3}k}{(2\pi)^{3}}(\vec{k}^{2}%
+m^{2})^{1/2}e^{i\vec{k}\cdot(\vec{x}-\vec{x}^{\prime})}.
\]
Also, at any point $\vec{x}$ the expected value of the field vanishes and its
variance diverges,
\begin{equation}
\langle\phi(\vec{x})\rangle_{0}=0\quad\text{and}\quad\langle\phi^{2}(\vec
{x})\rangle_{0}=\frac{1}{2}\int\frac{d^{3}k}{(2\pi)^{3}}\frac{1}{(\vec{k}%
^{2}+m^{2})^{1/2}}~.
\end{equation}
Note, however, that in ED the divergent energies and variances are epistemic
notions, so that once \textquotedblleft\emph{we free ourselves from the
delusion that probabilities are physically real things\textquotedblright\ }we
see that nothing physical is actually diverging.
\newline

\noindent \textbf{Acknowledgements: }I am grateful to N\'{e}stor Caticha, D. Bartolomeo, C. Cafaro, A. Giffin, P. Goyal, D. T. Johnson, K. Knuth, S. Nawaz, M. Reginatto, and C. Rodr\'{\i}guez, for many useful discussions on entropic dynamics.

\end{document}